\documentclass[preprint]{emulateapj}
\slugcomment{Accepted for Publication in the Astrophysical Journal}
\shortauthors{Zaritsky et al.}
\shorttitle{}

\begin{document}
\title{Evidence for Two Distinct Stellar Initial Mass Functions : Probing for Clues to the Dichotomy}
  
\author{Dennis Zaritsky}
\affil{Steward Observatory, University of Arizona, 933 North Cherry Avenue, Tucson, AZ 85721}

\author{Janet E. Colucci}
\affil{Carnegie Observatories, 813 Santa Barbara St., Pasadena, CA, 91101 }

\author{Peter M. Pessev}
\affil{Gemini South Observatory, c/o AURA Inc., Casilla 603, La Serena, Chile}

\author{Rebecca A. Bernstein}
\affil{Carnegie Observatories, 813 Santa Barbara St., Pasadena, CA, 91101}

\author{Rupali Chandar}
\affil{Department of Physics and Astronomy, The University of Toledo, 2801 West Bancroft Street, Toledo, OH, 43606}

\email{dzaritsky@as.arizona.edu}

\begin{abstract} 
We present new measurements of the velocity dispersions of eleven Local Group globular clusters using spatially integrated spectra, to expand our sample of clusters with precise integrated-light velocity dispersions to 29, over 4 different host galaxies. This sample allows us to further our investigation of the stellar mass function among clusters, with a particular emphasis on a search for the driver of the apparent bimodal nature of the inferred stellar initial mass function. We confirm our previous result that clusters fall into two classes. If, as we argue, this behavior reflects a variation in the stellar initial mass function, the cause of that variation is not clear. The variations do not correlate with formation epoch as quantified by age, metallicity quantified by $[ {\rm Fe/H}] $, host galaxy, or internal structure as quantified by velocity dispersion, physical size, relaxation time, or luminosity. The stellar mass-to-light ratios, $\Upsilon_*$, of the high and low $\Upsilon_*$ cluster populations are well-matched to those found in recent studies of early and late type galaxies, respectively. 
 
\end{abstract}


\section{Introduction}
\label{sec:intro}

The stellar initial mass function (IMF) is integral to a wide range of topics in extragalactic  astronomy. For the sake of simplicity, and having lacked definitive evidence to the contrary, it has been generally assumed that the IMF is universal. However, the exact functional form remains weakly constrained. Uncertainties in derived quantities that depend on the IMF are reminiscent of those ubiquitous decades ago in quantities that depend on the Hubble parameter. Different investigators then freely applied a ``correction" factor in transforming between one preference of H$_0$ and another, made explicit by the now endangered use of $h$. Similar uncertainties are now quoted when translating between ``Salpeter", ``Kroupa", and ``Chabrier" IMFs \citep{salpeter,kroupa01,chabrier}. However, with the IMF, unlike with the Hubble constant, we are not only dealing with a global normalization problem, but potentially with variations that vary within stellar systems, from system to system, and across time. If the adopted IMF is incorrect, grave consequences potentially follow for the ensuing physical interpretation. On a number of different fronts, it is appearing likely that at least the lower mass portion of the IMF varies significantly across stellar populations.

This is the third in our series of papers \citep[][hereafter Papers I and II, respectively]{z12,z13} that explores the nature of the IMF using dynamically measured masses of local group star clusters to infer stellar mass-to-light ratios, $\Upsilon_*$, as a function of age, metallicty, parent galaxy, and any other property of clusters that can be measured. 
To compare the stellar populations of the clusters, we use simple stellar evolution models to convert the current value of $\Upsilon_*$ for each cluster to that it would have at an age of 10 Gyr. This quantity is expressed as $\Upsilon_{*,10}$. In Paper I, we found that our small initial sample consisted of clusters that grouped around two different values of $\Upsilon_{*,10}$. After some discussion, we concluded with our preferred interpretation that this finding reflected the existence of different stellar initial mass functions, with one set of clusters having an IMF more consistent with the Salpeter function, $dN/dm \propto m^{-2.35}$,  and the other with IMFs in which the power law turns over at sub-solar masses \citep[for example,][]{kroupa01,chabrier}. This result joined a growing set of studies that identify variations in the lower end of the mass function among galaxies in the local universe \citep[such as][]{treu10,vandokkum,spiniello,cappellari12}, but provides the unique advantage of providing such measurements for much simpler, dark-matter free populations that are nearly of a single age and metallicity, potentially enabling us to track the cause of those variations.

In Paper II, we focused on exploring the potential deleterious effect of two-body relaxation on our measurements of the IMF. As clusters evolve, low mass stars 
can be ejected from the cluster, leading to values of $\Upsilon_{*,10}$ that are lower than would be calculated from exclusively modeling the stellar evolution of the initial stellar population. As such, we were concerned that the group of older clusters in our initial sample, which lie in the low $\Upsilon_{*,10}$ population, were affected by this phenomenon. Although we did address this concern in Paper I using published dynamical modeling of clusters, our conclusion was sensitive to ill-determined adopted parameters. In Paper II, we demonstrated that relaxation was indeed not a significant concern for clusters in our sample by increasing the mass range of the old clusters in our sample. Because relaxation affects low mass clusters more than high mass clusters, the lack of any trend with mass in the measured $\Upsilon_{*,10}$ values demonstrated empirically that relaxation is not responsible for the measured low $\Upsilon_{*,10}$ values in some of our clusters.

In this paper we continue our study by extending the sample in several important directions. We specifically target old clusters with high chemical abundance to provide a sample of old clusters that have chemical properties that overlap those of the younger clusters already in our sample. We also specifically target the two old clusters (NGC 2257 and NGC 6535) for which existing velocity dispersion measurements suggest large values of $\Upsilon_{*,10}$, and therefore bottom heavy IMFs \citep{mclaughlin,z12}. We aim to confirm, and refine, the velocity dispersion measurements of these two important, potentially unique, clusters. Finally, we target a set of clusters of intermediate age ($9 < \log({\rm age) }  < 10$, age in years) to populate this important region in parameter space. We briefly describe the data and data reduction in \S2, although it follows very closely what was done in Papers I and II. We present our results, combine with our previous published results, and provide our interpretation of the results in \S3. We summarize our findings in \S4.
   
\section{The Data}

Our spectroscopic data consist of observations obtained with the Magellan Clay telescope (6.5m) and the MIKE high resolution spectrograph \citep{mike} during June and August 2012 and October 2013. As in the previous papers in the series, our parent sample 
is the cluster compilation of \cite{mclaughlin}, from which we also draw the necessary ancillary data, such as age, half light radius, luminosity, and metallicity quantified by $ [{\rm Fe/H}]$. When using their model results, we choose those obtained using the Wilson models \citep{wilson}, which provide superior fits in the radial surface brightness profile in cases where distinctions can be made \citep{mclaughlin}.

We apply the spectroscopic drift scan technique described by \cite{colucci1} and used in our previous papers. We set the telescope in motion to raster scan the slit across the central region of the cluster ($\sim$ half light radius) during the exposure, defining both the angular length and height of the raster (both set to the same number).
The echelle slit is 0.75\arcsec x 5\arcsec. In taking multiple exposures on the alt-az telescope with no image rotator, each exposure covers a slightly different central area of the cluster and allows for cosmic ray removal. Each scan exposure is typically 1800 sec, for a combined total exposure that depends on the number of scans. These final exposure times are listed in Table \ref{tab:clusters}, along with other pertinent information.
The spectra have a wavelength coverage of approximately 3700 to 9800 \AA, with declining sensitivity and spectral resolution toward the blue end and significant background contamination toward the red.  Our analysis therefore focuses on the region between 5500 and 7000\AA\ where we have excellent data and a large number of suitable absorption features. We reduce the spectra using the MIKE IDL pipeline (http://www.ucolick.org/$\sim$xavier/IDL/index.html). Further details of the observations and data reduction can be found in \cite{colucci1} and in Paper I.

%%% the wavelength range of the spectra need to be checked

The spectra are not uniformly of the same high quality as in our previous studies even with comparable total exposure times (between 3000 and 40,000 seconds; see Table \ref{tab:clusters}) because  we targeted lower luminosity, lower surface brightness clusters to expand the cluster parameter range. We will discuss the impact of this choice when we discuss the uncertainties in our velocity dispersions and final estimates of $\Upsilon_*$. However, one clear failure is that for one of our key target clusters, NGC 2257, we were unable to obtain a velocity dispersion measurement at all and so this target is not included in Table \ref{tab:clusters}. Our experience here suggests that we have reached the end of the available clusters of certain types, such as intermediate ages, for this type of investigation given current capabilities. 

\subsection{Measuring Velocity Dispersions}

A discussion motivating our approach for measuring line-of-sight velocity dispersion, $\sigma_V$, is presented in Paper I, and we briefly review it here.
We fit a Gaussian broadening function to each of the same set of absorption lines selected in Papers I and II. We reject any lines in the object spectra that are either clearly blends or suffer some other complication and, after fitting, we reject lines that do not produce an acceptable fit, where acceptable is defined by $\chi_\nu^2 < 2.3$ over the spectral breadth of the line. To calculate $\chi_\nu^2$ we adopt a per pixel uncertainty determined from the fluctuations about a flat continuum in line-free areas of the spectrum, which we set to be the same for the full spectral range. Our results are not highly sensitive to this value because we only use these $\chi_\nu^2$ values to remove questionable fits from further consideration, and visual inspection confirms that those lines that have been rejected by this criteria are indeed poorly fit.

Once the line fitting is complete, we calculate the overall best value of $\sigma_V$ using the same approach as in Papers I and II.  We calculate a weighted mean of the $\sigma_V$ measurements from the individual spectral line fits. We proportionally increase the uncertainties associated with those individual fits so that the value of $\chi_{\nu}$ about the mean is 1. 
A key feature of our approach is that the factors by which we rescale the upward and downward uncertainties are different.
We do this because a number of potential systematic issues,
blended lines, poor continuum fit, focus errors, all bias the apparent line broadenings upwards. We chose in Paper I to 
scale the downward uncertainties to be 3 times larger than the upward ones, and we continue the same approach here. The net effect of this scaling is to diminish the influence of lines that have $\sigma_V$ larger than the mean.
We calculate the uncertainty on our final cluster $\sigma_V$, by identifying the range of $\sigma_V$'s that result in  $\chi^2 - \chi^2_{min} < 2.71$, corresponding to the 90\% confidence interval (see Table \ref{tab:clusters}).  This is a rather complicated process of which one might be somewhat suspicious, but the final uncertainties in the $\sigma_V$'s are eventually validated by comparing the empirical scatter in the derived $\Upsilon_{*,10}$ values within each of the two populations to the calculated uncertainties. 

We reference our values of $\sigma_V$ to the results from Paper I by analyzing standard stars in common to each of the observing runs using the same technique that is used to measure $\sigma_V$ for the clusters (the Paper II results were done with the Las Campanas 100 inch and are therefore on an independent, but apparently consistent, system, see below). We then remove in quadrature the excess ``velocity dispersion" found ($\sim 2.3$ km sec$^{-1}$ for the three runs we are currently presenting relative to the data presented in Paper I). The resulting correction to $\sigma_V$ is in all cases here smaller than the statistical uncertainty found using the technique described above, but it does illustrate that the ultimate systematic limit of this instrument/telescope combination is reached for systems with $\sigma_V \sim 3$ km s$^{-1}$.  For a more complete description and discussion of the issues involved we refer the reader to Paper I.

One cluster (NGC 6388) was observed with the Las Campanas 100-inch and included in the sample presented in Paper II and reobserved with the MIKE spectrograph on Magellan as part of this work. Our initial velocity dispersion measurement was 17.99$^{+0.46}_{-0.52}$ km s$^{-1}$ (Paper II) and our current measurement is 18.71$^{+0.51}_{-0.57}$ km s$^{-1}$. The difference is 0.72$^{+0.67}_{-0.77}$ km s$^{-1}$ and entirely consistent with zero. For this cluster, we average the measurements and adopt a velocity dispersion of 18.35$^{+0.49}_{-0.55}$ km s$^{-1}$. 

\begin{figure}[htbp]
\plotone{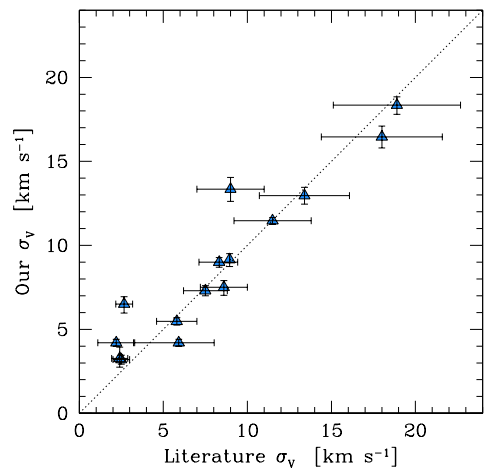}
\caption{A comparison between independent, previously published measurements of the cluster velocity dispersions and our measurements. The literature values are generally obtained from radial velocity measurements of individual member stars. The comparison includes all clusters for this study and Papers I and II for which we were able to find literature values. The line represents the 1:1 relation and is not a fit. The nature of the outliers is discussed in the text, but statistically the agreement is as expected. The principal advantage that our measurements bring to this type of work is the improved precision necessary to uncover subtle differences in $\Upsilon_*$.}
\label{fig:sigcomp}
\end{figure}

Further validation of the velocity dispersion measurements is drawn from Figure 1, where we compare to literature values when available. 
We obtain previous measurements from the compilation of \cite{mclaughlin}, which includes literature values of $\sigma_V$ for 11 of our clusters (over the entire age range), and other sources that include 4 more \citep{meylan,fischer,lane,origlia}. As we noted in Paper II,  for NGC 362 \cite{fischer} do not present a value of the dispersion so we calculate the standard deviation of the individual stellar velocities they provide, after excluding the sources they identified as either non-members or binary stars. As noted in Paper I, the one outlier at small velocity dispersion (NGC 1866), has a literature velocity dispersion that is strongly influenced by the acceptance/rejection of a single star \citep{lane}. Finally, 
the one outlier at large velocity dispersion comes from a study by \cite{origlia} which also presents a velocity dispersion measurement for NGC 6441 ($10\pm2$ km s$^{-1}$) that disagrees significantly not only with our measurement but with the value cataloged by \cite{mclaughlin}. The latter agrees with our measurement. For NGC 6441 we compare in the Figure with the $\sigma_V$ provided by \cite{mclaughlin} rather than that in \cite{origlia}, but for NGC 6440 we have found no alternative in the literature.
This extended external comparison is positive in that 10 of 15 (67\%) of the measurements agree to 1$\sigma$ and 
13 of 15 (87\%) agree to 2$\sigma$ (Figure \ref{fig:sigcomp}). 

In conclusion, Figure 1 demonstrates that the gain obtained in our measurement of the velocity dispersions of these systems is not one of accuracy, the literature values are clearly accurate, but rather one of precision. This precision is essential in unambiguously distinguishing the smaller $\Upsilon_{*,10}$ cluster population, for which there are many literature measurements available (see Figure 10 of Paper I), from the larger $\Upsilon_{*,10}$ population. Specifically, it is unclear from the literature data alone whether NGC 2257 and NGC 6353 are two unusual old clusters that belong in the larger $\Upsilon_{*,10}$ population. 

\section{Determining Masses and $\Upsilon_{*,10}$}

In Paper I we applied the prescription presented and tested by \cite{walker} to derive masses for spheroidal galaxies. As discussed previously in more detail \citep{z12a}, our finding that their prescription, when expressed in the formalism of galaxy scaling relations,
\begin{equation}
\log r_h = 2\log \sigma_V - \log I_h - \log \Upsilon_h - 0.73,
\label{eq:walker}
\end{equation}
where $r_h$, $I_h$, and $\Upsilon_h$ are the half light radius, the surface brightness within that radius, and the mass-to-light ratio within that radius, almost exactly matches the empirical finding of a scaling relation that applies to all stellar systems from stellar clusters to massive ellipticals
\begin{equation}
\log r_h = 2\log \sigma_V - \log I_h - \log \Upsilon_h - 0.75.
\label{eq:fm}
\end{equation}
\citep{z08,z11} provides additional confidence in this method. 
The empirical results verify that there is little scatter ($\sim 0.1$ dex) about this relationship for objects ranging from globular clusters to massive elliptical galaxies. We also demonstrated in Paper I that this method results in values of $\Upsilon_h$ in good agreement with published values calculated using more detailed dynamical models \citep{mclaughlin}.
Using Equation \ref{eq:fm}, we evaluate $\Upsilon_h$ for the 11 clusters of this study and present the results in Table \ref{tab:clusters}. 
For systems without dark matter, which we presume includes these clusters, $\Upsilon_* \equiv \Upsilon_h$. All photometric quantities are presented for the $V$ band.
The uncertainties in $\Upsilon_*$ are calculated using only the uncertainty in $\sigma_V$, as the internal uncertainties on the other parameters are proportionally much smaller. 

The scanning method is particularly well suited for our approach. The spectra are luminosity weighted means for the cluster stellar population. The resulting dispersion is not then the measure of the dispersion of the brightest stars. Instead, our measured velocity dispersions are the appropriate match to the photometric scale measurements obtained from the surface brightness profiles, which again are luminosity weighted means. An important consideration in the application of mass estimators is that the quantities being used all originate from the same population of dynamical tracers. It is possible for there to be distinct populations, such as globular clusters and satellite galaxies around a giant galaxy, that have different spatial distributions and kinematic properties. However, the application of a mass estimator to either population should result in consistent estimates of the mass of the central galaxy, if one uses the corresponding spatial and kinematic measurements. A strength of this study is that the velocity and radial scale come from the same stellar population, that which dominates the optical light, and so we do not require different populations to share the same spatial and kinematic characteristics.

Comparing values of $\Upsilon_*$ requires a method with which to calibrate out differences in $\Upsilon_*$ that are due to age or metallicity. One can compare (i.e. take the ratio of) the measured value of $\Upsilon_*$ to that calculated for a model stellar population, for example one with a Salpeter IMF, that has the same age and metallicity as the observed cluster. By examining if $\Upsilon_*/\Upsilon_{Salpeter}$ correlates with galaxy parameters, for example with galaxy mass or metallicity, one can begin to try to uncover the cause of the variations. This is what is now typically done in studies of galaxies \citep{conroy,cappellari12}. We take a slightly different approach, although the aim is similar. We identified a simple graphical method to ``renormalize" the values of $\Upsilon_*$ by noting in Paper I that the evolutionary tracks for models of single populations of log(age) $>8$ are approximately parallel lines in the space of log(age)$-$log($\Upsilon_*$). We therefore slide each cluster up or down such a line (slope = 0.64, that is calibrated to intercept the observed $\Upsilon_*$ at the age of the specific cluster) to the point at which log(age) $= 10$ and refer to this value of the mass-to-light ratio as $\Upsilon_{*,10}$. The slope we use here for the evolution is slightly shallower than used in Papers I and II, leading to most of the upper $\Upsilon_{*,10}$ population having slightly lower values of $\Upsilon_{*,10}$ than in Papers I and II, but none of our conclusions here or in the previous papers are affected by this change. The systematic errors introduced by ignoring the small deviations from a straight line in the evolutionary models are smaller than the statistical uncertainty in $\Upsilon_*$ (see Figure 9 in Paper I). 

\subsection{Adding to the Sample of Old Systems}

\label{sec:old}

We specifically selected 
our new set of clusters to address several perceived shortcomings of the original sample. We grew the number of old ($> 10$ Gyr old) clusters to explore a larger range of $[ \rm{Fe/H} ]$ so as to examine whether metallicity plays a role in the apparent dichotomy among cluster $\Upsilon_{*,10}$ values. The results of such a test can be thought of in two different ways. First, metallicity could give rise to a systematic error, for example through a template mismatch error in the derivations of $\sigma_V$. 
By finding clusters that populate both the high and low $\Upsilon_{*,10}$ populations and yet have similar metallicities, the likelihood of a systematic error in $\Upsilon_{*,10}$ due to metallicity diminishes. 
On the other hand, if certain values of $\Upsilon_{*,10}$ correspond exclusively to specific metallicity ranges, and we have no reason to suspect a systematic error, then we might have found evidence for mass function variations that depend on metallicity.

\begin{figure}[htbp]
\plotone{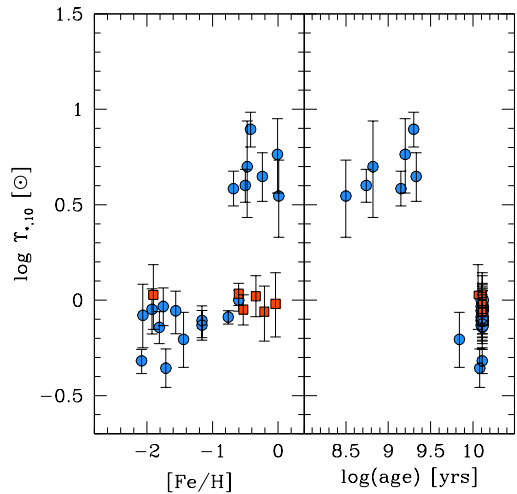}
\caption{$\Upsilon_{*,10}$ vs. $[ {\rm Fe/H} ]$ and age. Clusters from Papers I and II are plotted as blue circles, while a subsample of clusters presented here (see text) are plotted as red squares. The two cluster populations now clearly overlap in $[{\rm Fe/H} ]$, demonstrating that metallicity alone is not responsible for the two cluster populations.}
\label{fig:clustersz}
\end{figure} 

In Figure \ref{fig:clustersz} we add to the existing data from Papers I and II the subsample of old clusters in the new sample that populate the low $\Upsilon_{*,10}$ branch. With those new data, we now see that for over a factor of 2 in $[ {\rm Fe/H} ]$ there is no indication that metallicity affects the values of $\Upsilon_{*,10}$ for the clusters in the lower branch. Furthermore, there is now significant overlap in $[ {\rm Fe/H} ]$ among clusters in both branches of $\Upsilon_{*,10}$. These results indicate that metallicity, as measured by $[ {\rm Fe/H} ]$, does not play a leading  role either in distorting the values of $\Upsilon_{*,10}$ nor in producing different IMFs.

\begin{figure}[htbp]
\plotone{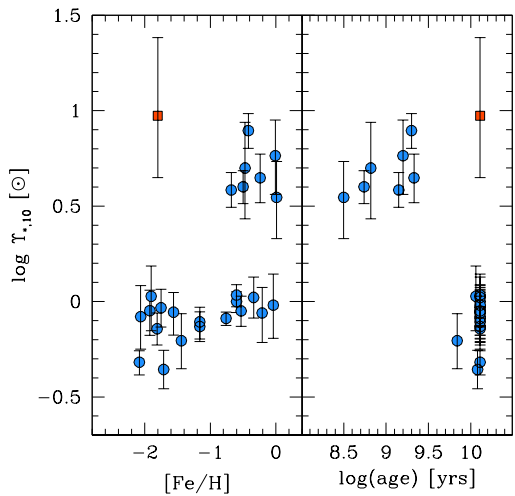}
\caption{$\Upsilon_{*,10}$ vs. cluster $[ {\rm Fe/H} ]$ and age. Clusters from Papers I, II, and those already presented in Figure \ref{fig:clustersz} are plotted as blue circles, while NGC 6535 is plotted as a red square. The addition of NGC 6535 has provide at least one old cluster that belongs to the population of large $\Upsilon_{*,10}$ clusters, demonstrating that age alone is not responsible for the two cluster populations.}
\label{fig:oldcluster}
\end{figure}

In addition to selecting clusters that explore a larger metallicity range, we also want to determine if there are any old clusters with large $\Upsilon_{*,10}$. The existence of such clusters would demonstrate that the large values of $\Upsilon_{*,10}$ found for intermediate age clusters in our previous work was not the result of age-related systematic errors in understanding either the dynamics or luminosities of those clusters. We targeted two clusters for observation, NGC 2257 and NGC 6535, because the existing literature values of $\sigma_V$ for these suggested that they might belong to the high $\Upsilon_{*,10}$ population. As we have mentioned before, we were unable to obtain a useful spectrum for NGC 2257, and only obtained one for NGC 6535 after spending over 46000 seconds of exposure time. The result for NGC 6535 is shown in Figure \ref{fig:oldcluster} and confirms the membership of this cluster in the upper $\Upsilon_{*,10}$ population. NGC 2257 remains an intriguing target. Its $\sigma_V$ literature value (5.8 km s$^{-1}$)  is sufficiently large that we suspect on the basis of the results in Figure \ref{fig:sigcomp} that it is unlikely to be incorrect by more than 2 km s$^{-1}$. Even if this cluster's velocity dispersion is ultimately as low as 3.8 km s$^{-1}$, the cluster would still lie on the upper $\Upsilon_{*,10}$ branch. 

The result for NGC 6535 warrants additional discussion due to its importance as the sole confirmed old cluster in the upper $\Upsilon_{*,10}$ branch. There are two potential sources of concern regarding its velocity dispersion (these are concerns in general, but more critical for NGC 6535 because it is currently a solitary cluster in that region of parameter space). The first concern is that  the velocity dispersion is contaminated by foreground/background stars, particularly because the cluster's Galactic latitude is 10$^\circ$. Our drift scanning method minimizes the influence of a small number of bright stars, whether they are cluster members or not, because a small fraction of the total exposure time is spent on those stars. However, for this cluster we are fortunate to have another argument as well against the role of contamination. \cite{pryor} present a velocity dispersion measurement based on high-resolution spectra of 26 stars. As important as the acceptable agreement between our $\sigma_V$ measurements ($3.27 ^{+0.67}_{-0.53}$ vs. $2.4\pm 0.45$ km sec$^{-1}$ for our and theirs respectively) is their measurement of the mean radial velocity, $-245$ km sec$^{-1}$. For that velocity they would have been able to easily remove Galactic contamination. Even if one attributes the slightly larger than 1$\sigma$ discrepancy between our measurement and theirs to contamination, the resulting $\Upsilon_{*,10}$ still places the cluster in the upper $\Upsilon_{*,10}$  branch. The second concern, related to its low Galactic latitude and small distance (3.9 kpc), is that dynamical interactions such as bulge or disk shocking, are artificially inflating the velocity dispersion. Simulations \citep{anders} show that such interactions tend to remove low mass stars from clusters and lead to low $\Upsilon_*$. One confirmed case, where star counts identify a deficiency of low mass stars, NGC 6366 \citep{paust09}, does not have an inflated velocity dispersion and our analysis confirms its low $\Upsilon_*$ \citep{z12}. Furthermore, we note that $\sigma_V$ for NGC 6535 would need to inflated to a fairly specific value in this scenario. If $\sigma_V$ had been inflated to 5 km sec $^{-1}$ rather than our measured value of $3.27$, we would calculate $\Upsilon_* \sim$ 25, extreme even for the high $\Upsilon_{*,10}$ branch. However, it is difficult to argue based on such circumstantial evidence that we have not inadvertently found the one cluster whose velocity dispersion is currently, moderately inflated. Because of this unknown, and other dynamical oddities that one could imagine, we need to either increase the number of old clusters with large $\Upsilon_{*,10}$ or, even better, confirm the presence of low mass stars through star counts, as described below.

\begin{figure}[htbp]
\plotone{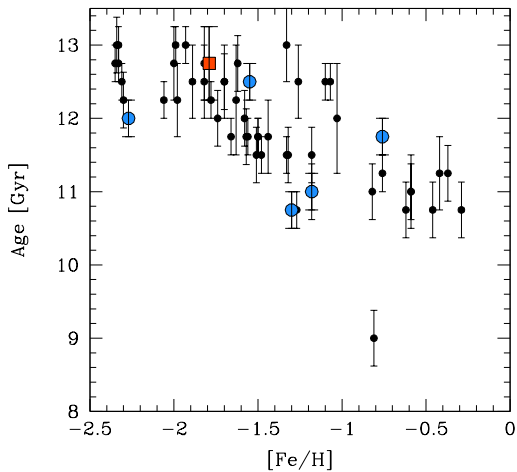}
%figure30
\caption{$[ {\rm Fe/H} ]$ vs. age for the sample of \cite{vdb} and those clusters in our sample that are in common. The red square denotes NGC 6535, the one old Galacitc clusters found so far to have high $\Upsilon_{*,10}$. The data split into two sequences, the upper is associated with clusters formed in the Galactic disk and the lower with accreted clusters by \cite{leaman}.}
\label{fig:vdb}
\end{figure}

If one accepts that NGC 6535 is not in an unusual dynamical state, then the result for NGC 6535 addresses several issues. First, as mentioned, it decreases the likelihood that an age-related systematic error accounts for the large values of $\Upsilon_{*,10}$
for the intermediate age clusters. Second, because NGC 6535 is a Galactic cluster, it removes the possibility that the high $\Upsilon_{*,10}$ values are limited to the Magellanic Clouds. It is possible that NGC 6535 was tidally stripped from a low mass galaxies, but that galaxy is unlikely to be either Magellanic Cloud because those are thought to be on their initial approach to the Galaxy \citep{besla}. We explore this possibility further using the two Galactic cluster populations identified by \cite{leaman}. They found that Galactic globular clusters fall into two populations in the age-metallicity parameter space, which they attribute to clusters that formed in the disk vs. those that formed in satellites and were accreted into the halo. In Figure \ref{fig:vdb} we show the position of the six clusters that we have in common with \cite{vdb} and \cite{leaman}. Despite having only six clusters in common, one of those is NGC 6535. NGC 6535 is not isolated in terms of the disk/halo populations so we cannot use this approach to identify it as an accreted cluster. If the \cite{leaman} interpretation is correct, we conclude that the distinction between disk and accreted cluster is not related to the $\Upsilon_{*,10}$ dichotomy, because high $\Upsilon_{*,10}$ clusters are found in both branches of their diagram. This conclusion can also be reached by noting that the LMC contains clusters in both the high and low $\Upsilon_{*,10}$ populations and will eventually be accreted by the Galaxy.
Third, NGC 6535 is a putative bottom-heavy cluster that is significantly closer than those in the Magellanic Clouds ($\ge 50$ kpc), providing at least one such cluster for which high resolution imaging should be able to probe the lower part of the main sequence (see \S 4.2 for more discussion on this aspect).

\subsection{Intermediate Age Clusters}

\begin{figure}[htbp]
\plotone{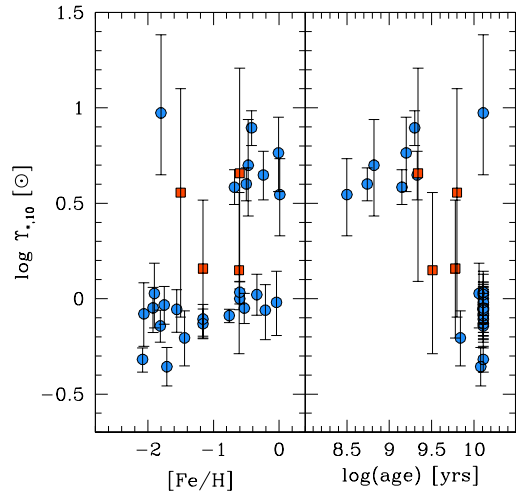}
\caption{$\Upsilon_{*,10}$ vs. $ [{\rm Fe/H} ]$ and age. Clusters from Papers I, II, and those already presented in Figures \ref{fig:clustersz}  and \ref{fig:oldcluster} are plotted as blue circles, while the subsample of intermediate age clusters discussed in \S is plotted as red squares. These additional four clusters have sufficiently large uncertainties that we are unable to place them unambiguously in either population, or even perhaps in between. The velocity dispersions are uncertain both because the clusters are of lower surface brightness, and hence the spectra have lower S/N, and because the clusters tend to have low (few) km sec$^{-1}$ velocity dispersions, which are difficult to measure precisely with this instrument.}
\label{fig:intclust}
\end{figure}

The four new clusters in our sample with $\log({\rm age}) < 10$ are unfortunately of low surface brightness and produced poor S/N spectra. Furthermore, these four clusters turn out to have small velocities dispersions $(< 4$ km s$^{-1})$ which puts them in a regime where they are more susceptible to systematic errors. As such they not only have larger error bars in Figure \ref{fig:intclust}, but potential systematic errors similar to the size of the error bars. It is therefore futile to discuss whether the clusters lie in either population or between the two populations. These four clusters may be better targets for individual star spectroscopy, but again their low velocity dispersions suggest that a higher velocity resolution spectrograph than typically available at 6-10m class telescopes may be needed.

\section{Discussion}

\subsection{The Specter of Systematic Errors}

With the current sample, we have eliminated age and metallicity as both potential sources of systematic error in the determination of $\Upsilon_{*,10}$ and as physical drivers, or critical parameters, related to variations in the IMF. Additionally, certain causes of concern, such as binaries, now require finely tuned scenarios. For example, if binaries are responsible for the high $\Upsilon_{*,10}$ values, we would require binaries to dominate at intermediate ages, so as to preferentially distort the measurement of $\sigma_V$ for these systems, and yet still  affect at least one old cluster (NGC 6535). The tightness of $\Upsilon_{*,10}$ within each of the two populations argues against a stochastic binary contribution. Instead, our principal remaining concern for a systematic error lies with the mass estimator. It is conceivable that there are two physically distinct forms of clusters, but with the same IMF, for which different numerical constants are needed in Eqs. 1 and 2. 

\begin{figure}[htbp]
\plotone{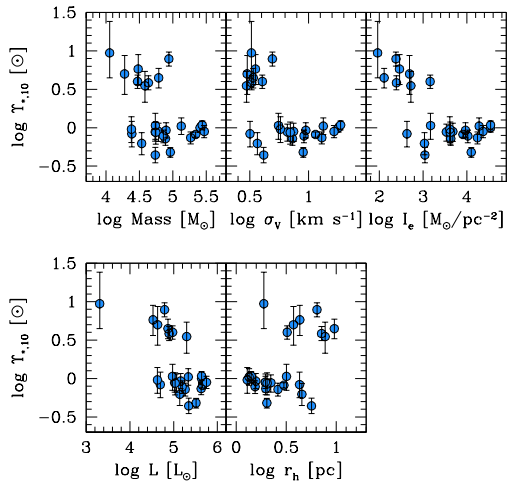}
%figure10
\caption{$\Upsilon_*$ vs. empirical cluster parameters. All clusters, with the exception of the four poorly constrained intermediate age clusters (NGC 339, NGC 2121, NGC 2193, KRON 3) are included here. In all panels, the clusters from each of the two populations overlap in parameter space and when considering either class alone, there is no significant trend between $\Upsilon_{*,10}$ with any of the plotted measured parameter. These results argue against a dependence of the mass estimator on one of the measured structural characteristics.}
\label{fig:empirical}
\end{figure}

A potential signpost of such a problem would be if the two $\Upsilon_{*,10}$ populations also clearly divide into two distinct categories, for example high and low velocity dispersions or high and low surface densities. Of course, such a dichotomy could also help explain differences in the IMF, but a connection with one of the parameters that is directly involved in our calculation of $\Upsilon_*$ would be worrisome. In Figure \ref{fig:empirical} we show the values of $\Upsilon_{*,10}$ relative to all of the relevant empirical parameters of our study ($\sigma_V$, $I_e$, $r_h$, $L$)  and the derived one, $M$. In all cases there is overlap between the two populations of clusters, thereby ruling out each of these parameters as uniquely defining a $\Upsilon_{*,10}$-class of clusters. This result does not remove all concern about possible systematic errors of this type, because there could still be differences in the mass normalization of the estimator due to a hidden parameter, such as the formation timescale of the cluster. 

\begin{figure}[htbp]
\plotone{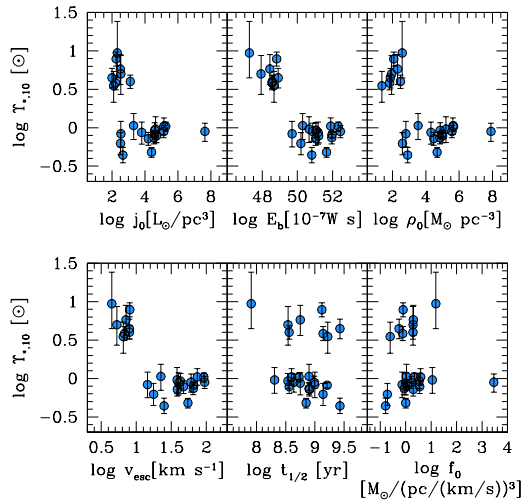}
%figure13
\caption{$\Upsilon_*$ vs. model cluster parameters (the model parameters are based on adopted values of a single $\Upsilon_*$ from stellar population modeling using a universal IMF). All clusters, with the exception of the four poorly constrained intermediate age clusters (NGC 339, NGC 2121, NGC 2193, KRON 3) are included here. The only parameters that show no overlap between the two populations are $v_{esc}$ and $E_b$. See text for interpretation and description of parameters.}
\label{fig:modelclust}
\end{figure}

Beyond the directly empirical quantities examined above, we now use the compendium by \cite{mclaughlin} to examine the relation between $\Upsilon_{*,10}$ and various model-derived cluster parameters (Figure \ref{fig:modelclust}).  It is critical to note here that these derived quantities depend on the cluster mass, which in this case is calculated using standard stellar population models to obtain a mass-to-light ratio and the luminosity. Because those authors adopt a single, universal IMF in the calculations, these comparisons provide null tests. If the cluster IMF is actually universal, as assumed in this modeling, what differences do we find among our apparent high and low $\Upsilon_{*,10}$ populations? 

We see in Figure \ref{fig:modelclust} that certain quantities, namely the half-mass relaxation time, $t_{1/2}$ and the central phase space density, $f_0$, overlap almost completely between the two types of clusters. These results, particularly the $t_{1/2}$ result, argue against internal dynamical evolution as being critical in artificially producing two populations (providing further confirmation of our results in Paper II). Other parameters, namely the central luminosity or mass densities, $j_0$ and $\rho_0$, show more limited overlap, but still offer examples of systems with the same value of each of those properties that lie in either of the two $\Upsilon_{*,10}$ populations. Finally, and most intriguing, is the separation of the two cluster populations with the escape speed, $v_{esc}$ and binding energy, $E_b$. 

\begin{figure}[htbp]
\plotone{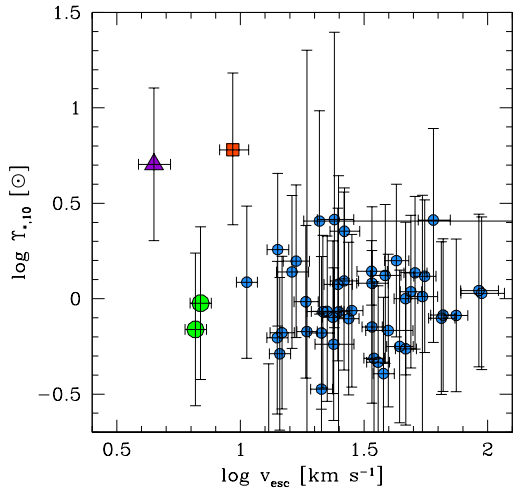}
%figure14
\caption{$\Upsilon_*$ vs. $v_{esc}$ for clusters in the compilation of \cite{mclaughlin}. The purple triangle circle represents NGC 6535, the red square NGC 2257 and the green circles NGC 5053 and 5466. Although the uncertainties in $\Upsilon_{*,10}$ are larger than what we obtain from our measurements of $\sigma_V$, this literature sample enables us to enlarge the sample and find two clusters that have both low values of $\Upsilon_{*,10}$ and $v_{esc}$, demonstrating that $v_{esc}$ does not define a sample of clusters for which the adopted mass estimator, Eq. 1, is invalid. }
\label{fig:models1}
\end{figure}

This finding is perhaps worrisome in terms of our dynamical modeling. Could the internal dynamics and structure of these clusters be sufficiently different that the mass estimator of Eq. 1 does not hold? To pursue this issue, we augment the sample shown in Figure \ref{fig:modelclust} by including  all of the clusters in \cite{mclaughlin} for which literature $\sigma_V$'s exist (Figure \ref{fig:models1}). Although the uncertainties are larger because of the larger uncertainties in $\sigma_V$, we see that for $v_{esc} < 1$ there are two additional clusters (NGC 5053 and 5466) that have low $\Upsilon_{*,10}$, in agreement with the bulk of the old Galactic clusters. We conclude that simply having a low $v_{esc}$, as inferred using the photometric mass estimates) does not in and of itself lead directly to large derived values of $\Upsilon_{*,10}$ when using Eq. 2.

If instead we now accept that our values for $\Upsilon_{*,10}$ are correct for the high $\Upsilon_{*,10}$ population, which are $\sim$ 0.6 dex larger than the values for the low $\Upsilon_{*,10}$ population, then $v_{esc}$, which scales with mass and hence $\Upsilon$, should similarly increase by about 0.6 dex, placing these clusters well in overlap with the low $\Upsilon_{*,10}$ clusters. Therefore, within the scenario that our values of  $\Upsilon_{*,10}$ are correct, our clusters are not unusual in their $v_{esc}$. Likewise, due to its $\propto M^2$ behavior, the clusters in the $E_b$ plot also move over to overlap the low $\Upsilon_{*,10}$ population. 

Ultimately, the only convincing way to address the potential for systematic errors in the mass estimator is to identify variations in the mass functions of these systems  using different observational approaches. Below we discuss the use of star counts, which currently does not resolve the issue, but has the potential to do so. We also compare our results to those being obtained for early and late-type galaxies, which show a similar dichotomy in inferred IMFs as our clusters. Because there is no reason to expect systematic errors in our method to produce $\Upsilon_*$ values that agree with those derived for galaxies using different techniques, we consider the agreement discussed in \S 4.3 to be additional circumstantial evidence in favor of our findings. 

\subsection{Star Counts}

The gold standard measure of the stellar mass function is direct star counts. Unfortunately, those efforts are complicated in the current context. Locally, within our galaxy near the Sun, the method is at its most robust and results, while showing some scatter, are roughly consistent with an IMF that turns over at low stellar masses \citep[$\sim 0.2-0.4 M_\odot$,][]{bastian}. Moving outward, to encompass star clusters similar to those in our study, the results are of limited value because Galactic clusters are almost entirely in the low $\Upsilon_{*,10}$ population according to what we have found so far. Indeed a state-of-the-art study of such clusters \citep{leigh} finds no evidence for variations in the IMF. As such, the few of our clusters for which star count measurements exist (NGC 104, 362, and 6093) all have relatively shallow slopes \citep[$<1.69$,][]{paust} indicative of bottom-light populations, in agreement with our results, but not surprisingly so.

Efforts to extend such work to the important Magellanic Cloud clusters is also limited, but for another reason. The \cite{glatt} sample has three clusters in common with our sample (NGC 121, NGC 339, and Kron 3) for which they measure the present day mass function power law slope index. However, because of the much greater distance to these clusters the counts only reach to masses as low as $\sim 0.5 M_\odot$, so they do not probe the stellar mass regime where the turnover in the mass function is expected and most prominent. Interestingly, there is one study of the stellar mass function in the Magellanic Clouds that does extend sufficiently faint \citep{kalirai}. That study avails itself of the extremely deep {\sl HST} imaging around 47 Tuc (121 orbits) to reach stars of $M \sim 0.37 M_\odot$ and concludes that while the mass function at low (sub solar) masses is somewhat shallower than that of a Salpeter mass function, it does not show a significant turnover down to this mass limit. This deep study of the SMC field is complemented by studies of lower mass dwarf galaxies \citep{geha} and although there is a suggestion of a trend in the character of the low mass end of the mass function with velocity dispersion, the differences among the SMC and less massive galaxies, where the data and analysis methods are similar, are not yet statistically significant. The differences become statistically significant only if one includes the MW and ellipticals, where the data and analysis methods to obtain the mass slopes are substantially different. Nevertheless, this is evidently a strong line of investigation that merits attention. 

Although the strengths of the star count approach are evident, it is not a panacea. The correction for binaries and other multiples, and assumptions regarding the mass ratios of such, can be quite large and, if those corrections are functions of stellar mass or other properties \citep{milone}, they will affect the inferred slopes. In essence the star counts are lower limits on the mass functions if conservative assumptions are made regarding companion stars. Furthermore, while a dynamical measurement is sensitive to all the mass, and therefore to remnants, the star count method must make assumptions regarding the upper end of the mass function.

Among our clusters, NGC 6535 stands out as a clear target for such work. It is unique so far among the Galactic clusters for resulting in a large $\Upsilon_{*,10}$. NGC 2257, which also appears to have a large $\Upsilon_{*,10}$, but which we were unable to confirm is also a tempting target. Unlike the Magellanic Cloud clusters that appear to be bottom heavy, NGC 6535 is much closer (6.8 kpc) and more amenable to study.  We noted in Paper I the interesting case of NGC 6366 as a proof-of-concept because it is found to be unusually bottom light both from star counts \citep{paust09} and from our dynamical analysis using the literature data from \citep{mclaughlin}. In this case the mass function anomaly is explained due to preferential loss of low mass stars given the position of the cluster in the Galaxy and its strong dynamical interaction with the bulge \citep{paust09}.

\subsection{Comparison to Galaxies}

\begin{figure}[htbp]
\plotone{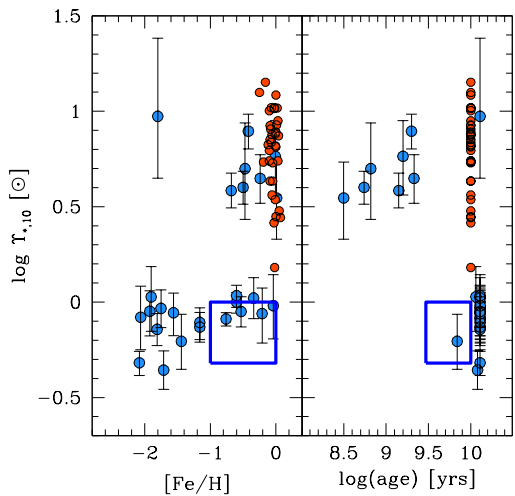}
%figure20
\caption{$\Upsilon_*$ for clusters and galaxies. All clusters, with the exception of the four poorly constrained intermediate age clusters (NGC 339, NGC 2121, NGC 2193, KRON 3) are included here and values of $\Upsilon_{*,10}$ are plotted. For galaxies we include two results. First, the red circles show the early type data from \cite{conroy} corrected to V-band as described in the text. Second, the blue box denotes the mean value of $\Upsilon_*$ for the disks of disk galaxies adopted from the dynamical survey of \cite{martinsson}. The width of the box in either panel is notional and does not represent a measurement of an age or metallicity spread. Adopting an extremely low age of 3 Gyr to correct the measured value of 
$\Upsilon_*$ to $\Upsilon_{*,10}$ defines the upper boundary of the box, whereas assuming an age of 10 Gyr defines the lower boundary. Because both ages are extreme, we expect  disks to lie somewhere within the box.}
\label{fig:galaxies}
\end{figure}

We now explore how our two population of clusters compare to the range of $\Upsilon_*$ found among galaxies. These comparisons are complicated by the compound populations present in galaxies, so we cannot directly compare values at 10 Gyr. For early type galaxies, which presumably have negligible young stellar populations this omission may not be severe. We select the data from \cite{conroy} for our comparison. We convert from their R band values of $\Upsilon_*$ to V-band by assuming a single color for all of their early types \citep[V$-$R$_C = 0.8$;][]{fukugita} and adopting (V$-$R)$_\odot = 0.4$. Those galaxies are then plotted in comparison to our clusters in Figure \ref{fig:galaxies}.

For the late type galaxies, and in particular for the disks of late type galaxies, we adopt the average value of $\Upsilon_*$ reported by \cite{martinsson} who use measurements of the vertical velocity dispersion in disks to constrain the mass of the disk itself, separate from the dark matter contribution. This is the most direct dynamical measurement of $\Upsilon_*$ so we favor it over alternate estimates. For the comparison to our data we have also had to convert from their value of $\Upsilon_*$ in the K-band (0.31 $\pm 0.07$), adopting a representative disk color of V$-$K$= 3.1$ \citep[for Sc galaxies;][]{aaronson} and (V$-$K)$_\odot = 1.65$, resulting in an estimate of the V-band $\Upsilon_*$ of $0.48$. However, this value is not $\Upsilon_{*,10}$, which is almost certainly larger depending on the star formation history that is appropriate for disks. If, for example, we assume that the disks have a single age population with  an age of 3 Gyr (log(age) = 9.5), almost certainly too small for the mean stellar age, and apply our correction to arrive at $\Upsilon_{*,10}$, we calculate $\Upsilon_{*,10} = 1$. Presumably the appropriate value for comparison to our sample lies between $0.48$ and 1, well in the range of $\Upsilon_{*,10}$ of the lower branch of the cluster population. We conclude that the early type galaxy stellar populations are consistent with those of the clusters in the upper population and that the stellar populations in the disks of late type galaxies are consistent with those of the clusters in the lower one. 
The cluster results therefore provide completely independent, but consistent, evidence for variations in the IMF at the magnitude level seen between early and late type galaxies. 

\section{Conclusions}

Because the possibility of a non-universal IMF is important to a wide range of astronomical questions, we obtained more data to enlarge our sample of clusters with precise velocity dispersions and further test our original claims. In contrast to the increasing independent evidence for IMF variations in galaxies
\citep{vandokkum,cappellari12,geha}, the clusters provide a simple, single population view into this phenomenon. Here we presented data for eleven additional clusters in the Milky Way, the Magellanic Clouds, and the Fornax dwarf galaxy, bringing the total sample to 29, with which to explore aspects of the problem. 

First, we specifically set out to measure the stellar mass-to-light ratio, $\Upsilon_*$, of metal-rich old clusters, a class that was absent in our previous papers. We found that those clusters exhibit the same behavior as the previously studied old clusters, thereby demonstrating that metallicity, as quantified by $[{\rm Fe/H}]$, is not the driver for the apparent or real variations in $\Upsilon_*$.  The old clusters now clearly show, over a range in mass and metallicity, remarkably uniform values of $\Upsilon_*$, demonstrating that various potential sources of ``noise" in this study (orbital parameters, orbital history, internal relaxation) are negligible in relation to the detected effect.

Second, we specifically set out to measure $\Upsilon_*$ for two candidate clusters with old ages and suspected high $\Upsilon_*$'s, based on literature values of $\sigma_V$. If confirmed, either of these would be the only such clusters in our sample. We were able to measure $\sigma_V$ for one of these two (NGC 6535) and confirmed the literature value of $\sigma_V$ that alerted us to this cluster in the first place, although with the improved precision that enables us to differentiate it more significantly from the low $\Upsilon_*$ population. We failed to measure $\sigma_V$ for the second (NGC 2257). With the existence of at least one old cluster with high $\Upsilon_{*,10}$, we show that age (or dynamical effects) alone are not responsible for the dichotomy in cluster properties. The existence of this cluster also rules out intermediate age binary stars as the sole cause of high $\Upsilon_{*,10}$ values. 

Third, we specifically set out to increase the number of intermediate age ($9 < $ log(age) $< 10$) clusters in our sample. This effort was less successful in the sense that our measured velocity dispersions are not of sufficient precision to provide any additional insight. The problem is two-fold. First, to expand our sample required us to observe clusters that are both less luminous and of lower surface brightness than previously studied clusters, making those observations that much more difficult. Additionally, these are lower mass clusters in general and so their velocity dispersions are low, and close to the systematic limits of the current instrumental capabilities. Both of these aspects worked against us in this instance. Nevertheless, these are key clusters to investigate and should be the focus of future studies. 

We have now robustly demonstrated the existence of two classes of stellar cluster in the Local Group. We argue against various systematic errors and in favor of the hypothesis that values of $\Upsilon_*$ reflect variations in the IMF. While the empirical result is robust, the interpretative one is more speculative. Nevertheless, the mean values of $\Upsilon_*$ for each of the high and low $\Upsilon_*$ populations matches precisely the values obtained in recent work on early and late type galaxies, respectively. This agreement provides circumstantial support for our interpretation. The clusters are key in that they should provide a simpler to interpret entry into the phenomenon of IMF variations. Nevertheless, we have yet to identify a characteristic property that maps uniquely onto the putative IMF variations. The cause of IMF variations lies beyond the obvious physical characteristics of these clusters.

\bigskip
{\it Facility}: \facility{Magellan: Clay (MIKE Spectrograph)}

\begin{acknowledgments}
DZ acknowledges financial support from 
NSF grant AST-1311326 and NASA ADAP NNX12AE27G and thanks NYU/CCPP for its hospitality during his various visits there. 
R. C. acknowledges support from NSF through CAREER award 0847467. 

\end{acknowledgments}

\begin{deluxetable*}{lllrrrrrrrr}
\tablewidth{0pt}
\tablecaption{Stellar Cluster Data}
\tablehead{
\colhead{NGC} &
\colhead{Host} &
\colhead{Tel.} &
\colhead{$t_{exp}$} &
\colhead{log(age)} &
\colhead{log(L$_{V}$)}&
\colhead{$[{\rm Fe/H}]$} &
\colhead{log $r_h$} &
\colhead{log(I$_h$)} &
\colhead{$\sigma_V$} &
\colhead{$\Upsilon_{*}$} \\
&&&[s]&[Gyr]&[L$_{\odot}$]&&[pc]&$[L_\odot pc^{-2}]$&km s$^{-1}$&$[\odot]$\\
}
\startdata
NGC 339 & SMC &  Mag & 23400 &  9.80 &  4.88 &$-$1.50 & 1.070  & 1.94 & $3.93^{+1.07}_{-1.28}$&  2.68$^{+1.93}_{-2.30}$ \\
NGC 2121 & LMC &  Mag & 12600 &  9.51 & 5.08  &$-$0.61 & 1.050  & 2.18 & $ 2.56^{+0.52}_{-0.56}$&  0.69$^{+0.79}_{-0.86}$ \\
NGC 2193 & LMC &  Mag &  10800&  9.34 & 4.36 &$-$0.60 & 0.666 & 2.23 & $2.76^{+0.76}_{-0.78}$&  1.72$^{+1.61}_{-1.65}$  \\
NGC 6388 & MW &  Mag & 3600&  10.11 & 5.64 &$-$0.60 & 0.150  & 4.54 & $18.35^{+0.49}_{-0.55}$&1.41$^{+0.14}_{-0.15}$\\
NGC 6440 & MW &  Mag &  25200  & 10.11 & 5.34 &$-$0.34 & 0.129 & 4.28 & $13.34^{+0.71}_{-0.72}$&  1.23$^{+0.27}_{-0.27}$ \\
NGC 6441 & MW &  Mag &  3600  & 10.11 & 5.75 &$-$0.53 & 0.292 & 4.37 & $16.46^{+0.64}_{-0.66}$&  1.05$^{+0.18}_{-0.19}$ \\
NGC 6528 & MW &  Mag & 10800  & 10.11 & 4.63 &$-$0.04 & 0.111 & 3.61 & $5.77^{+0.47}_{-0.50}$&  1.12$^{+0.39}_{-0.42}$ \\
NGC 6535 & MW &  Mag & 46200 & 10.11 & 3.31 &$-$1.80 & 0.276 & 1.96 & $3.27^{+0.67}_{-0.53}$& 11.06$^{+2.68}_{-2.12}$ \\
NGC 6553 & MW &  Mag & 2700  & 10.11 & 5.04 &$-$0.21 & 0.309 & 3.62 & $6.99^{+0.47}_{-0.54}$&  1.02$^{+0.31}_{-0.36}$ \\
FORNAX 4 & FORNAX &  Mag & 7800 & 10.06 & 4.97 &$-$1.90 & 0.503 & 3.17 & $5.55^{+0.44}_{-0.50}$&  1.16$^{+0.39}_{-0.44}$ \\
KRON 3 & SMC &  Mag & 46200 & 9.78 & 5.08 &$-$1.16 & 1.039  & 2.20 & $3.19^{+0.57}_{-0.58}$&  1.04$^{+0.84}_{-0.85}$ \\

\enddata
\label{tab:clusters}
\end{deluxetable*}

\end{document}